\documentclass[twocolumn,preprintnumbers,superscriptaddress,reprint,apl]{revtex4}

\usepackage{graphicx}
\usepackage{bm}
\usepackage{amsmath}
\usepackage{amssymb}
\renewcommand{\vec}[1]{\bm{#1}}

\begin{document}


\title{Hybrid reflections from multiple x-ray scattering in epitaxial bismuth telluride topological insulator films}

\author{S\'ergio L. Morelh\~ao}
\affiliation{Department of Physics, University of Guelph, Guelph, Ontario N1G\,1W2, Canada}
\address{Institute of Physics, University of S\~ao Paulo, S\~ao Paulo 05508-090, Brazil}
\author{Stefan Kycia}
\affiliation{Department of Physics, University of Guelph, Guelph, Ontario N1G\,1W2, Canada}
\author{Samuel Netzke}
\affiliation{Department of Physics, University of Guelph, Guelph, Ontario N1G\,1W2, Canada}
\author{Celso I. Fornari}
\affiliation{National Institute for Space Research, S\~ao Jos\'e dos Campos, S\~ao Paulo 12227-010, Brazil}
\author{Paulo H. O. Rappl}
\affiliation{National Institute for Space Research, S\~ao Jos\'e dos Campos, S\~ao Paulo 12227-010, Brazil}
\author{Eduardo Abramof}
\affiliation{National Institute for Space Research, S\~ao Jos\'e dos Campos, S\~ao Paulo 12227-010, Brazil}

\date{\today}

\begin{abstract}
Epitaxial films of bismuth telluride topological insulators have received increasing attention due to potential applications in spintronic and quantum computation. One of the most important properties of epitaxial films is the presence of interface defects due to lateral lattice mismatch since electrically active defects can drastically compromise device performance. By describing hybrid reflections in hexagonal bismuth telluride films on cubic substrates, in-plane lattice mismatches were characterized with accuracy at least 20 times better than using other X-ray diffraction methods, providing clear evidence of 0.007\% lateral lattice mismatch, consistent with stress relaxation associated with van der Waals gaps in the film structure.
\end{abstract}


\maketitle

Semiconductor and optical industries make broad usage of thin epitaxial films and multilayers. X-ray diffraction (XRD) is an essential tool for the characterization of epitaxial thin films, giving the necessary information about composition, orientation, and structural perfection for both quality control and development of new device technologies. Structure refinement by XRD simulation is the most used approach for accurate data analysis, which is only as reliable as the quality of the best achieved fit of the data. To this purpose, all features (peaks in intensity) in a scan of intensity must be identified \cite{ehs17}. Any disagreement between experimental and simulated data call into question the actual structure of the thin film \cite{slm02a}.

Hybrid reflections (HRs) from multiple x-ray diffraction are known to cause extra features in diffraction data of thin films. In general, they can be easily avoided or identified by changing the sample azimuth \cite{ehs17}, or they can be used as a tool for studying heteroexpitaxial films \cite{ish81,slm93a,slm93b,jzd16,yzz16} and superlattices \cite{slm02c}. In a new class of epitaxial systems with potential applications in spintronic and quantum computation, film and substrate materials have quite different lattices \cite{che10,wan11,lee12,hoe14}. In these cases, HRs can be nearly inevitable in any scan of intensity. There are no available approaches to describe their occurrence neither to use them for analyzing one of the most important properties of expitaxial systems that seriously impact the final performance of the devices, which is the amount of defects due to lateral lattice mismatch at the film/substrate interface.

Bismuth telluride (Bi$_2$Te$_3$) has been recently established as an archetype for the three-dimensional topological insulators \cite{zha09,che09}. Since intrinsic conduction through only topological surface states can be obtained in high-quality thin films \cite{wan11,hoe14}, there has been significant investigation on the growth parameters of thin films by molecular beam epitaxy (MBE) where epitaxial films of Bi$_2$Te$_3$ with hexagonal lattice grow along the $c$ axis on cubic (111) substrates  \cite{yyl10,ste14,cif16a,bon17}. van der Waals bonds along film thickness allow epitaxy on substrates of large lattice mismatch \cite{kom99,guo15,lhe13,gha17,lit17}, and the effects of substrate choice on film quality, surface morphology, and mobility of charge carriers are also subjects of recent investigations \cite{kam15,gin16,wag18}. In this work, we provide a general description of hybrid reflections in such awkward systems, giving the necessary equations to identify and use HRs in the most common XRD methods employed worldwide for studying epitaxial films. Experimentally, we use an in-house x-ray diffractometer to demonstrate and exploit HRs in bismuth telluride films grown by MBE on BaF$_2$ (111), which has been so far the most suitable substrate for these films due to the very small in-plane lattice mismatch of only 0.04\% \cite{ste14,cif16a}.

HRs arise from the differences between two distinct crystal lattices that share a common interface. They are produced by sequences of successive ${\rm h}_s {\rm k}_s {\rm l}_s$ reflections in the substrate and ${\rm h}_f {\rm k}_f {\rm l}_f$ reflections in the film, in principle, without any specific order or number of reflections involved on each possible sequence. In reciprocal space, hybrid reflections have diffraction vectors
\begin{equation}\label{eq:qstar}
    \vec{Q}^*=\sum_{f,s} (\vec{Q}_f + \vec{Q}_s)
\end{equation}
where $\vec{Q}_f$ and $\vec{Q}_s$ stand for diffraction vectors on either film or substrate lattices, respectively. In real space, their scattering direction obey the same rule of any diffraction vector, making an angle
\begin{equation}\label{eq:tthQstar}
\Theta=2\arcsin(Q^*/Q_{\rm max})
\end{equation}
(usually called $2\theta$) with the incident x-ray direction, $Q_{\rm max}=4\pi/\lambda$ for x-rays of wavelength $\lambda$. Regarding the outward normal direction $\hat{\vec{n}}$ to the common interface, the hybrid diffraction vectors can be decomposed in terms of perpendicular $Q^*_{\bot}=\vec{Q}^*\cdot\hat{\vec{n}}$ and in-plane $\vec{Q}^*_{\|}=\vec{Q}^* - Q^*_{\bot}\hat{\vec{n}}$ components. Then, the x-ray incidence angle at the interface falls within the interval
\begin{equation}\label{eq:thinc}
\theta_i = \Theta/2 \pm \arcsin(Q^*_{\|}/Q^*)\,,
\end{equation}
depending on which azimuth $\vec{Q}^*$ is excited.

The most relevant fact about HRs is that their occurance is a direct consequence of differences between both lattices. In other words, when film and substrate share identical unit cells and crystallographic orientations, as in homoepitaxy, there will be no extra reciprocal lattice nodes since all $\vec{Q}^*$ vectors fall on top of the usual ones from both lattices. Then, the multi-scattering process implied by Eq.~(\ref{eq:qstar}) symplifies to the well known multiple diffraction of X-ray in single crystals \cite{slc10,slm02b}.

Films coherently strained to their substrates are the simplest systems where HRs can be observed \cite{ehs17}. Along the specular truncation rod, $Q^*_{\|}=0$, $\theta_i=\Theta/2$, and in the scan of intensity versus $2\theta$, i.e. the typical $\theta$-$2\theta$ scans as usually called, hybrid peaks show up around scattering angles given by Eq.~(\ref{eq:tthQstar}) where $\vec{Q}^*$ has only the $Q^*_{\bot}$ component, i.e. $\Theta=2\arcsin(Q^*_{\bot}/Q_{\rm max})$. It is worthwhile to reemphasize that observation of hybrid peaks depends on sample azimuth, and the axial divergence of the x-rays plays an important role in observing these peaks. Wider axial divergence results in increased chance of exciting HRs unintentionally.

Formation of interface defects is the typical mechanism by which the elastic stress due to in-plane lattice mismatch relaxes \cite{tan17,yao18}. When the film is relaxed, film and substrate no longer share exactly the same in-plane lattice parameter, and $Q^*_{\|}\neq0$ even for the HRs that would be along the specular truncation rod in non-relaxed films. Consequently, a $\theta$-$2\theta$ scan with narrow angular acceptance in $2\theta$ may not even capture those HRs that are slightly off the substrate truncation rods \cite{slm07}. Such HRs can be seen by placing the detector at $2\theta=\Theta$, Eq.~(\ref{eq:tthQstar}), and carrying out standard rocking curve measurements (scan of intensity versus incidence angle $\theta$), wide enough to satisfy the incidence angle in Eq.~(\ref{eq:thinc}).

Measuring off-specular-rod HRs as a function of the incidence angle has been the most reliable procedure to evidence small amount of interface defects in epitaxial films, with accuracy better than the 0.05\% limit of standard methods via asymmetric reflections \cite{jzd16}. Due to the very small mismatch of 0.04\%, relaxation of thin Bi$_2$Te$_3$ films on BaF$_2$ (111) is beyond the detectability limit of current methods. Here, a detailed description on how to select suitable HRs to characterize the coherence of film/substrate interfaces is provided, which can also be usefull when investigating topological insulator films on substrates of larger lattice mismatch \cite{lhe13,wag18}.

\begin{figure}
\includegraphics[width=3.4in]{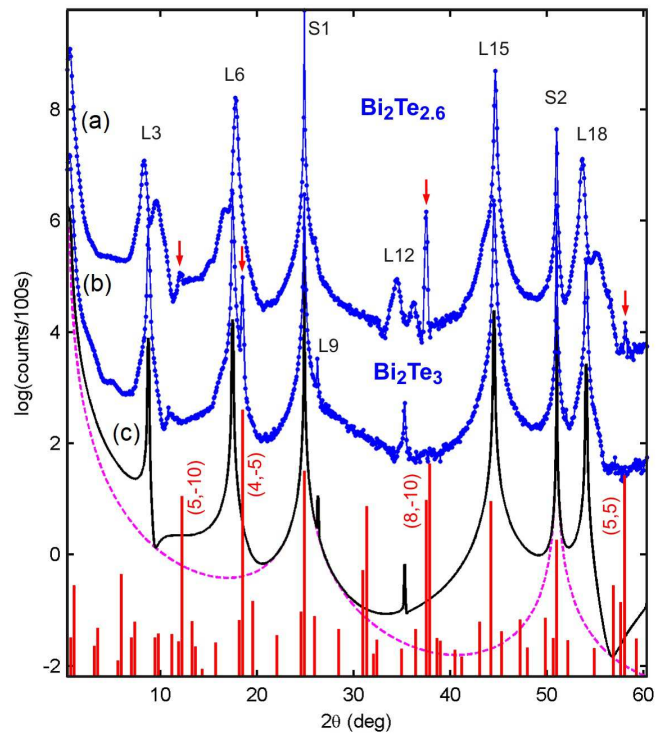}\\
  \caption{Long-range $\theta$-$2\theta$-scans in Bi$_2$Te$_{3-\delta}$ films on BaF$_2$ (111) substrates. (a,b) Experimental scans in films with (a) Te deficit of $\delta=0.4$, and (b) no Te deficit, $\delta=0$. (c) Simulated scans in film/substrate (solid line) and substrate (dashed line) \cite{slm17}. L3, L6,$\ldots$L18 are reflections from the film, while S1 and S2 are the 111 and 222 reflections of the substrate. Visible HRs are pointed out by arrows and labeled according to their $(n,m)$ indexes in Eq.~(\ref{eq:tthnm}). Position of possible HR peaks are indicated by vertical (red) lines.}\label{fig:tthscans}
\end{figure}

Bismuth telluride films have been grown on freshly cleaved (111) BaF$_2$ substrates using a Riber 32P MBE system that contains a nominal Bi$_2$Te$_3$ effusion cell and two extra Te sources \cite{cif16a,cif16b}. To compensate the loss of tellurium during growth, the ratio $\Phi_R$ between the beam equivalent pressures of Te sources and Bi$_2$Te$_3$ effusion cell can be adjusted from $\Phi_R = 0$ (no extra sources of Te) to about $\Phi_R = 6$. Substrate temperature $T_S$ and additional Te impact the deficit of Te in the films \cite{ste14,cif16a}. With $T_S=270^\circ$C and $\Phi_R = 3$, a high quality Bi$_2$Te$_3$ film with no Te deficit was obtained, while $T_S=290^\circ$C and $\Phi_R = 1$ led to a Bi$_2$Te$_{3-\delta}$ film with significant Te deficit $\delta\simeq0.4$. Both films were grown for 2 hours, at rates of 0.21\,\AA/s, resulting in thicknesses of 150\,nm. The lateral lattice parameter in bulk materials is known to increase with the deficit of Te, going from 0.4382\,nm for $\delta=0$ to about 0.4409\,nm for $\delta=0.4$  \cite{lin05,ste14}, i.e. a variation of 0.6\%.

X-ray measurements were carried out with a Huber four circle diffractometer sourced by a fine focus copper rotating anode configured with a double collimating multilayer optics followed by a double bounce Ge 220 channel cut monochromator. Bandwidth is 2\,eV for CuK$\alpha_1$ ($\lambda=1.540562$\,\AA). Electronic noise of the point detector is 0.08\,counts/s. Adjustment arcs of the goniometer head were used to align the $222_s$ substrate reflection with the $\varphi$ rotation axis of the diffractometer with an accuracy better than $0.01^\circ$. As reference for sample azimuth, the substrate $313_s$ and film $01\,20_f$ asymmetric reflections have been measured in co-planar diffraction geometry at the same azimuth \cite{cif16b}. Hence the in-plane direction $[110]$ of the film lattice coincides with the $[0\bar{1}1]$ direction of the substrate lattice, and we define $\varphi=0$ when these directions are in the horizontal diffraction plane pointing upstream. Positive rotation sense of the $\varphi$ axis is clockwise. The major difference of our x-ray diffractometer regarding commercial ones for thin film analysis is the narrow axial (vertical) divergence of about $0.005^\circ$, which is as small as the divergence in the horizontal diffraction plane.

\begin{figure}
\includegraphics[width=3.4in]{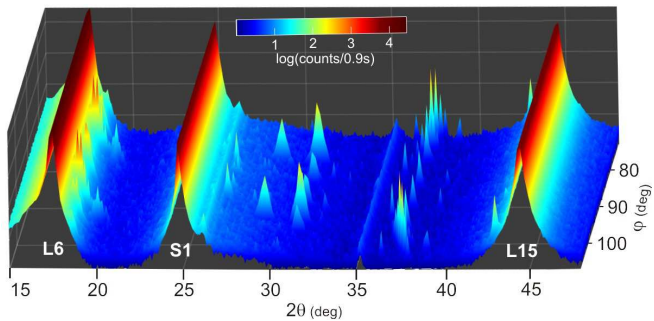}\\
  \caption{$\theta$-$2\theta$:$\varphi$ mesh scan in Bi$_2$Te$_3$/BaF$_2$(111), revealing some of the main HRs (peaks in $\varphi$) of this epitaxial system.}\label{fig:tthphimesh}
\end{figure}

Figs.~\ref{fig:tthscans}a,b show $\theta$-$2\theta$ scans along specular rods of the samples. The $00{\rm l}_f$ film reflections, labeled L3, L6,$\ldots$L18 (${\rm l}_f=3,6,\ldots,18$) are clearly visible, as well as the $111_s$ and $222_s$ substrate reflections, labeled S1 and S2, respectively. As demonstrated in previous publications \cite{ste14,slm17}, spliting of peaks L3, L6, L12, and L18 in the Bi$_2$Te$_{2.6}$ film (Fig.~\ref{fig:tthscans}a) are caused by the Te deficit that favours bilayers of Bi to form inside the van der Waals gap between two consecutive Bi$_2$Te$_3$ quintuple layers (QLs). The Te deficit in this film means that on average there is one bilayer of Bi for every 6.5 QLs (about 22 bilayers in the whole film thickness). For the second sample, the diffraction pattern in Fig.~\ref{fig:tthscans}b aggrees very well with the simulated pattern in Fig.~\ref{fig:tthscans}c for the Bi$_2$Te$_{3}$ film. In both experimental scans, there are several additional peaks (pointed out by arrows), in disagreement with the expected ones for this epitaxial system. These peaks are attributed to HRs.

In the case of this hexagonal-on-cubic growth, the outward normal direction $\hat{\vec{n}}$ has different representation in each lattice, it stands for the [001] and [111] directions in the film and substrate lattices, respectively. It implies that all HRs have  $Q^*_{\bot}=\vec{Q}^*\cdot\hat{\vec{n}}=2\pi(n/a_0\sqrt{3}+m/c)$ and, as far as in-plane mismatch is negligible, HRs along the specular rod have scattering angles
\begin{equation}\label{eq:tthnm}
    \Theta_{nm} = 2\arcsin\left [\frac{\lambda}{2}\left (\frac{n}{a_0\sqrt{3}}+\frac{m}{c}\right )\right ]
\end{equation}
where $n=\sum_s({\rm h}_s+{\rm k}_s+{\rm l}_s)>0$ and $m=\sum_f{\rm l}_f$. By using $a_0=6.2001$\,\AA\, as the cubic lattice paremeter of BaF$_2$ and $c=30.497$\,\AA\, as the hexagonal lattice parameter along the film [001] direction, the $2\theta$ angles of all possible HRs along the specular rod where calculated and compared to experimental ones in Fig.~\ref{fig:tthscans}; details about this calculation as well as on the relative intensities of the $(n,m)$ families of HRs are given in the supplementary material.

Due to the large value of the $c$ parameter, HRs in this epitaxial system can be found in seemingly random location in standard $\theta$-$2\theta$ scans. Even in a diffraction system with narrow axial divergence, it is difficult to find an azimuth to perform a $\theta$-$2\theta$ scan without exciting a few HRs, as shown in the $\theta$-$2\theta$:$\varphi$ mesh scan in Fig.~\ref{fig:tthphimesh}.

\begin{figure}
\includegraphics[width=3.4in]{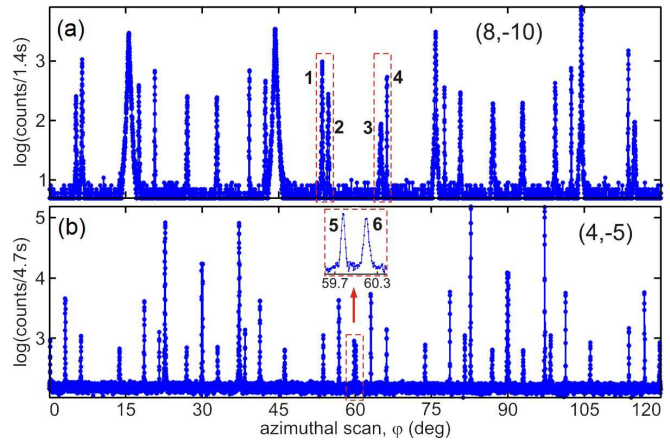}\\
  \caption{Azimuthal scans in Bi$_2$Te$_3$/BaF$_2$ at fixed incidence angles $\theta_i = \Theta_{nm}/2$, showing two families of ($n,m$) hybrid peaks. (a) $\theta_i = 18.74$ degrees, hybrids (8,-10), and (b) $\theta_i = 9.24$ degrees, hybrids (4,-5). HRs for characterizing lateral lattice mismatch are indicated by numbers.}\label{fig:phiscans}
\end{figure}

Families of ($n,m$) hybrid peaks follow the 3-fold axis symmetry of the growth direction. High resolution azimuthal scans of 120 degrees for the two most populated families are shown in Fig.~\ref{fig:phiscans}. Most hybrid peaks in these scans arise from simple sequences of only two reflections, one in the film and the other in substrate lattice or vice-versa, i.e. $\vec{Q}^*=\vec{Q}_f + \vec{Q}_s$ or $\vec{Q}^*=\vec{Q}_s + \vec{Q}_f$. Full indexation lists and a script on how to index these hybrids are given in the supplementary material. Here it is important to indentify what are the suitable cases that can be used for studying in-plane lattice mismatch in these films. Since the in-plane component of the diffraction vectors are easily described in terms of ${\rm h}_f{\rm k}_f$ indexes of the ${\rm h}_f{\rm k}_f{\rm l}_f$ reflections of the film hexagonal lattice, a convenient way to write down the in-plane component of $\vec{Q}^*$ as a function of the in-plane mismatch $\Delta a/a$ is, in a first order derivation,
\begin{equation}\label{eq:Qinstar}
    \vec{Q}^*_\| = -\vec{Q}_{f,\|}\,\Delta a/a
\end{equation}
where the in-plane component $\vec{Q}_{f,\|}$ of the film diffraction vector  is calculated for $\Delta a/a=0$.

For a pair of HRs where film and substrate reflections are equivalent but occuring on opposite sequences, $\vec{Q}^*_\|$ have the same magnitude for both HRs although with opposite signals. If such a pair is excitated at the same azimuth within the axial divergence, there will be a split of the observed hybrid peak as a function of the rocking curve angle, i.e., a split regarding the single peak seen for a sample with coherent strained film. In our diffractometer, the axial divergence is too narrow to excite more that one hybrid at the same azimuth. Then, we looked up for pairs of nearby hybrid peaks in the azimuthal scans where one of the peaks has a very small width, which in general indicates a hybrid with first reflection occurring in the substrate lattice, and hence, a possible pair with opposite sequences of equivalent reflections. The most suitable pairs we found are pointed out by numbers in the azimuthal scans in Fig.~\ref{fig:phiscans}. Peaks 1\&4: $\bar{2}2\,\bar{10}_f+044_s$ and $404_s+0\bar{2}\,\bar{10}_f$ is one pair; peaks 2\&3: $6\bar{2}4_s+1\bar{4}\,\bar{10}_f$ and $\bar{3}4\,\bar{10}_f+\bar{2}64_s$ is another pair; and peaks 5\&6: $4\bar{4}4_s+0\bar{4}\bar{5}_f$ and $\bar{4}4\bar{5}_f+\bar{4}44_s$ is the only pair where the peaks are less than 1 degree apart from each other.

\begin{figure}
\includegraphics[width=3.4in]{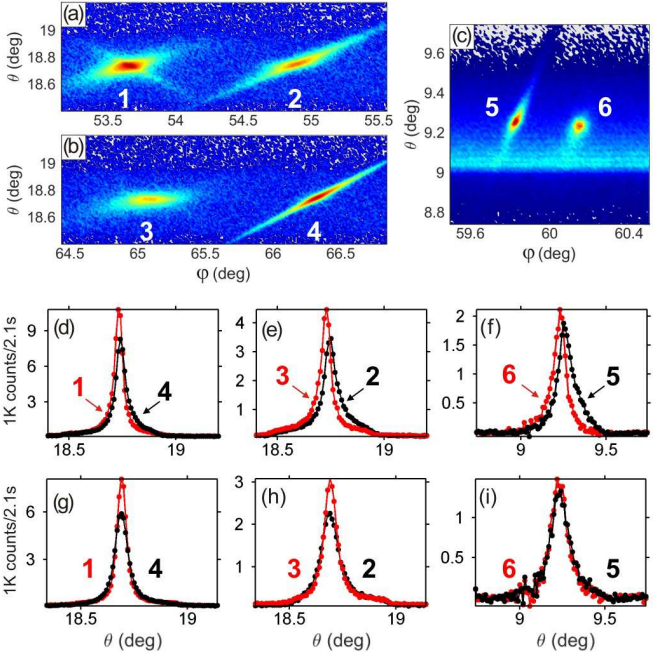}\\
  \caption{(a-c) Two-dimensional intensity profiles of the hybrid peaks highlighted in Fig.~\ref{fig:phiscans}. (d-i) Intensity of hybrid peaks as a function of the incident angle $\theta$ (rocking curves) for the samples with Bi$_2$Te$_3$ (d-f) and Bi$_2$Te$_{2.6}$ (g-i) films.}\label{fig:meshscans}
\end{figure}

To imitate a conventional x-ray diffractometer of wider axial divergence, of about 1 degree, we perform a $\theta$-$\varphi$ mesh scans around the hybrid peaks pointed out in Fig.~\ref{fig:phiscans}, and to compare their rocking curves we integrate the intensity in $\varphi$. For the Bi$_2$Te$_3$ film, these mesh scans are shown in Figs.~\ref{fig:meshscans}a-c, while the rocking curves of each pair are compared in Figs.~\ref{fig:meshscans}d-f. For the Bi$_2$Te$_{2.6}$ film, the rocking curves of the same pairs are compared in Figs.~\ref{fig:meshscans}g-i. Since hybrids of the (4,-5) family are at the  shoulder of reflection $006_f$ (peak L6 in Fig.~\ref{fig:tthscans}b), the contribution of this reflection have been subtracted from the rocking curves of peaks 5\&6 in Figs.~\ref{fig:meshscans}f-i.

It is interesting to note that the Bi$_2$Te$_3$ film is indeed relaxed, as expected in van der Waals epitaxy \cite{gha17}, while the Bi$_2$Te$_{2.6}$ film is perfectely matched to the substrate. The actual shift of hybrid peaks due to film relaxation is estimated from Eqs.~(\ref{eq:thinc}) and (\ref{eq:Qinstar}) by taking $\hat{\vec{k}_{\|}}$ as the in-plane direction of the incident wavevector $\vec{k}$ at the azimuth in which the hybrid is excited. Then, the shift in the rocking curve angle of hybrid peaks as a function of lateral lattice mismatch is given by
\begin{equation}\label{eq:thincxphi}
    \Delta\theta_i \simeq - \frac{\vec{Q}_{f,\|}\cdot\hat{\vec{k}_{\|}}}{Q^*}\frac{\Delta a}{a}\,.
\end{equation}
From the diffraction geometry of a single reflection, it is straighforward to conclude that $\vec{Q}_{f,\|}\cdot\hat{\vec{k}_{\|}}<0$ for all hybrids in which the first reflection takes place in the film, i.e. hybrids with diffraction vector $\vec{Q}^*=\vec{Q}_f + \vec{Q}_s$. Relaxation of the film implies in \mbox{$\Delta a/a=(a_{f,\|}-a_{s,\|})/a_{s,\|}<0$} when considering bulk values where $a_{f,\|}<a_{s,\|}=a_0/\sqrt{2}=4.3841$\,\AA. Therefore, in case of relaxation of the film, those hybrids with first reflection in the film such as hybrids 1, 3, and 6 shift towards smaller values of the incidence angle, $\Delta\theta_i<0$ in Eq.~(\ref{eq:thincxphi}), while hybrids 2, 4, and 5 shift towards higher angles, $\Delta\theta_i>0$, exactly as observed in the rocking curves in Figs.~\ref{fig:meshscans}d-f.

Values of peak shifting shown in Table~\ref{tab:1} for the Bi$_2$Te$_{3}$ film indicates a mismatch of $\Delta a/a=-7(\pm2)\times10^{-5}$, about 0.007\%, while for the Bi$_2$Te$_{2.6}$ film no mismatch could be detected. Variation in lattice mismatch due to fluctuation of room temperature is of the order of $1.1\times10^{-6}/K$. For the measured HR pairs with x-rays of 8\,keV, refraction corrections \cite{slm16} at the film/substrate interface can account for spliting the hybrid peaks by less than $0.0032^\circ$ (see supplementary material), which is smaller than our accuracy in measuring the split of rocking curve peaks. Therefore, misfits of 0.002\% in lateral lattice parameters is close to the minimum that can be detected by measuring HR pairs in this epitaxial system without extra protocols for temperature control better than a few degrees and corrections due to refraction. It is comparable to the most accuracte methods available for lattice parameter determination in single crystals \cite{wlb60,jaq13}.

In summary, by using a high resolution method for measuring lateral lattice mismatch in epitaxial films we have demonstrated that strain in the film lattice are easily relaxed through van der Waals gaps between monoatomic layers stacking along the growth direction, in spite of the very small mismatch. This is an important mechanism of elastic stress relaxation that can be harmless to device perfomance since the electric field around van der Waals bonds is much smaller than around covalent or ionic chemical bonds. In the film with Te deficit, one bilayer of bismuth in the van der Waals gap of every 6 or 7 quintuple layers of Bi$_2$Te$_3$ is capable of increasing van der Waals forces, stiffening the structure, and prevent the relaxation of the film. Otherwise, this film would relax to a larger lateral lattice parameter, of about 0.6\%, regarding the film without bilayers of bismuth.

\begin{table}[t]
\caption{In-plane mismatch $\Delta a/a$ of Bi$_2$Te$_{3}$ and Bi$_2$Te$_{2.6}$ films on BaF$_2$ (111), determined by measuring $\Delta\theta_i$ [Eq.~(\ref{eq:thincxphi})] from rocking curves of HR pairs in Figs.~\ref{fig:meshscans}d-i.}\label{tab:1}
\scriptsize{
\begin{tabular}{ccccccc}
  \hline\hline
   & & & \multicolumn{2}{c}{Bi$_2$Te$_{3}$} & \multicolumn{2}{c}{Bi$_2$Te$_{2.6}$} \\
   & & & \multicolumn{2}{c}{---------------------} & \multicolumn{2}{c}{---------------------} \\
   HR & $Q_{f,\|}$ &  & $2\Delta\theta_i$ & 2$\Delta a/a$ & 2$\Delta\theta_i$ & 2$\Delta a/a$ \\
   pair& (\AA$^{-1}$) & $\vec{Q}_{f,\|}\cdot\hat{\vec{k}_{\|}}/Q^*$ & $(^\circ)$ & $(\times10^{-4})$ & $(^\circ)$ & $(\times10^{-4})$ \\
  \hline
  1\&4 & 3.3098 & $\mp1.0176$  & 0.0086(35)  & -1.5(6)  & -0.0019(46)  & -0.3(8)\\
  2\&3 & 5.9667 & $\pm2.2352$  & 0.0195(43)  & -1.5(3)  & -0.0025(56)  & -0.2(4)\\
  5\&6 & 6.6195 & $\pm4.3824$  & 0.0299(58)  & -1.2(3)  &  0.0039(88)  &  0.2(4)\\
  \hline
  \hline
\end{tabular}}
\end{table}

See supplementary material for choice of reference frame, indexation of hybrid reflections, and corrections due to refraction.

The authors acknowledge CAPES (Grant No. 88881.119076/2016-01) and FAPESP (Grant No. 2016/22366-5) for financial support.


\clearpage

\large{SUPPLEMENTARY MATERIAL}

\section{reference frame}

Diffraction vectors $\vec{Q}={\rm h}\vec{a}^* + {\rm k}\vec{b}^* + {\rm l}\vec{c}^*$ are given in terms of the reciprocal unit cell vectors
$$\vec{a}^*=2\pi\dfrac{\vec{b}\times\vec{c}}{V_c},\, \vec{b}^*=2\pi\dfrac{\vec{c}\times\vec{a}}{V_c},\,{\rm and}\;
\vec{c}^*=2\pi\dfrac{\vec{a}\times\vec{b}}{V_c}$$
where $\vec{a}$, $\vec{b}$, and $\vec{c}$ are the unit cell vectors in real space and $V_c=(\vec{a}\times\vec{b})\cdot\vec{c}$ the unit cell volume.

To project the diffraction vectors of a crystal lattice onto a convenient $xyz$ reference frame used to describe intrumental $\theta$ and $\varphi$ angles, $\vec{A}=A_1\vec{a}^*+A_2\vec{b}^*+A_3\vec{c}^*$ is the reciprocal vector along the $\varphi$ rotation axis, and $\vec{B}=B_1\vec{a}^*+B_2\vec{b}^*+B_3\vec{c}^*$ is the direction in reciprocal space that lies in the incident plane (pointing upstream) when $\varphi=0$. Then, $$\hat{\vec{x}} = \hat{\vec{y}}\times\hat{\vec{z}},\, \hat{\vec{y}} = \hat{\vec{z}}\times\vec{B}/|\hat{\vec{z}}\times\vec{B}|,\,{\rm and}\;\hat{\vec{z}} = \vec{A}/|\vec{A}|\,.$$ In this frame, the incidente wavevector is written as $$\vec{k}=-\dfrac{2\pi}{\lambda}[\cos(\theta)\cos(\varphi)\hat{\vec{x}}+
\cos(\theta)\sin(\varphi)\hat{\vec{y}}+\sin(\theta)\hat{\vec{z}}]\,,$$ its in-plane direction as $$\hat{\vec{k}}_{\|}=-[\cos(\varphi)\hat{\vec{x}}+\sin(\varphi)\hat{\vec{y}}]\,,$$
and the diffraction vectors as
$$\vec{Q}^\prime=(\vec{Q}\cdot\hat{\vec{x}})\hat{\vec{x}}+
(\vec{Q}\cdot\hat{\vec{y}})\hat{\vec{y}}+
(\vec{Q}\cdot\hat{\vec{z}})\hat{\vec{z}}\,.$$

To describe hybrid reflections in the Bi$_2$Te$_3$/BaF$_2$(111) epitaxial system, we have used $A=[A_1\,A_2\,A_3]=[0\,0\,1]$ and $B=[B_1\,B_2\,B_3]=[1\,1\,0]$ for projecting the film diffraction vectors, whereas for the substrate diffraction vectors $A=[1\,1\,1]$ and $B=[0\,{-1}\,1]$ were used instead. As an example, consider the reflections $\bar{2}2\,\bar{10}_f$ of the film and $044_s$ of the substrate. Their respective diffraction vectors in the $xyz$ frame are $\vec{Q}^\prime_f = 3.3098\hat{\vec{y}}-2.0603\hat{\vec{z}}$ and $\vec{Q}^\prime_s = -3.3098\hat{\vec{y}}+4.6807\hat{\vec{z}}$, giving rise to the hybrid $\bar{2}2\,\bar{10}_f+044_s$ of diffraction vector $\vec{Q}^*=\vec{Q}^\prime_f+\vec{Q}^\prime_s=2.6204\hat{\vec{z}}$. It has no in-plane component as far as the lateral lattice parameter $a$ of the film matchs the substrate one, i.e.  $a=a_0/\sqrt{2}=4.3841$\,\AA.

\section{relative intensity of hybrid reflections in $2\theta/\theta$ scans}

\begin{figure*}
  \includegraphics[width=6.4in]{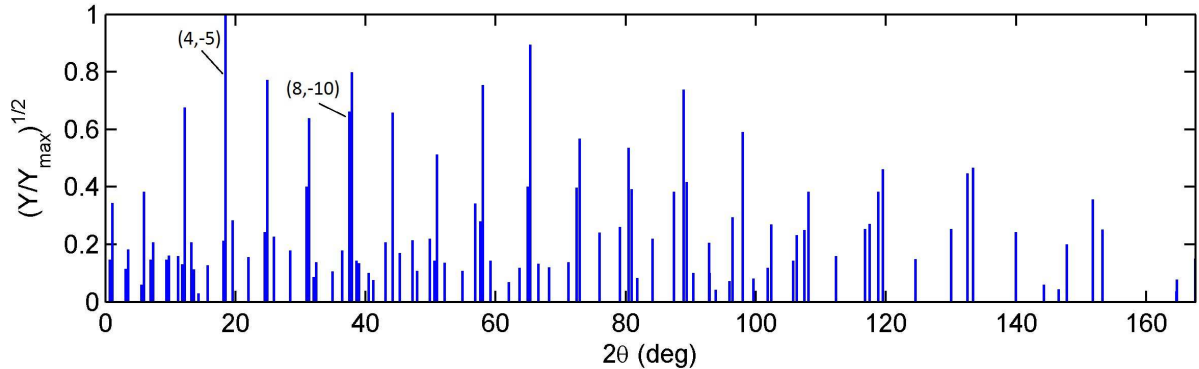}\\
  \caption{Calculated positions of main hybrid reflections along $2\theta/\theta$-scans in Bi$_2$Te$_3$ film stressed to BaF$_2$ (111) substrate. Relative intensities estimated by function $Y(2\theta)$.}\label{fig:mainhybrids}
\end{figure*}

With the purpose of simple identification of the main hybrid reflections, their relative intensities were compared on basis of products of film and substrate structure factor modules, $|F_f|$ and $|F_s|$ respectively. For a hybrid with $\vec{Q}^*=\vec{Q}_f+\vec{Q}_s$, its relative intensity is simply taken as proportional to $|F_f||F_s|$. Along $2\theta/\theta$ scans, the positions where hybrid peaks can be observed and their relative intensities are predicted by
\begin{equation}\label{eq:y}
Y(2\theta)=\sum_{f,s}|F_f||F_s|\delta(2\theta-\Theta_{nm})\nonumber
\end{equation}
where $f$ and $s$ run over all allowed film and substrate reflections of indexes ${\rm h}_f{\rm k}_f{\rm l}_f$ and ${\rm h}_s{\rm k}_s{\rm l}_s$, respectively.
$$\Theta_{nm} = 2\arcsin\left [\frac{\lambda}{2}\left (\frac{n}{a_0\sqrt{3}}+\frac{m}{c}\right )\right ]\,,$$
$n=\sum_s({\rm h}_s+{\rm k}_s+{\rm l}_s)>0$, and $m=\sum_f{\rm l}_f$. A plot of the $Y(2\theta)$ is shown in Fig.~\ref{fig:mainhybrids}. Note that this plot compares the relative intensities of the $(n,m)$ families of hybrid reflections, and not individual hybrids.

\section{Indexation of hybrid reflections in azimuthal scans}

A general and very friendly script in MatLab was written to index hybrid reflections in film/substrate epitaxial systems. As input, it requires the list of structure factors and lattice parameters of both film and substrate crystal lattices, as well as the instrumental $\theta$ and $\varphi$ angles where a hybrid peak is observed. Then, the script projects all film and substrate diffraction vectors onto the $xyz$ reference frame and searches in both lattices for the closest reflection in diffraction condition with the incident wavevector $\vec{k}$. Such search is carried out by minimizing the value of $\xi_1 = |k_1 - 2\pi/\lambda|$ where $\vec{k}_1=\vec{Q}^\prime_1 + \vec{k}$. When $\vec{Q}^\prime_1$ is a diffraction vector of the film lattice, it is also necessary that $\vec{k}_1\cdot\hat{\vec{z}}<0$. Otherwise, if $\vec{Q}^\prime_1$ is in the substrate lattice, then $\vec{k}_1\cdot\hat{\vec{z}}>0$. Once the 1st reflection is identified, the script searches for the 2nd reflection of diffraction vector $\vec{Q}^\prime_2$ in the other lattice by minimazing the value of $\xi_2 = |k_2 - 2\pi/\lambda|$ where $\vec{k}_2=\vec{Q}^\prime_2 + \vec{k}_1$. To be a valid hybrid of diffraction vector $\vec{Q}^*=\vec{Q}^\prime_1+\vec{Q}^\prime_2$ that can be observed in specular diffraction geometry, further restrictions apply: $\vec{Q}^*\cdot\hat{\vec{x}}=\vec{Q}^*\cdot\hat{\vec{y}}=0$ and their reflection indexes must provide the proper scattering angle $2\theta\simeq\Theta_{nm}$. The product $|F_f||F_s|$ of film and substrate structure factors may also be set as a criterion to validate hybrid reflections found by the script. Table~\ref{tab:1} shows all hybrids of families (8,-10) and (4,-5) identified with this script.

\begin{table*}[t]
\caption{Indexation of HRs measured in azimuthal scans with fixed incidence angle $\theta$. Peaks 1 to 24: $\theta=18.738^\circ$ [Fig. 3(a), main text]. Peaks 25 to 60: $\theta=9.243^\circ$ [Fig. 3(b), main text]. Experimental values of $\varphi$ have accuracy of about $0.02^\circ$, and values of $|F_f||F_s|$ have been normalized by a factor of $10^4$. }\label{tab:1}
\scriptsize{
\begin{tabular}{|c|c|c|c|c|c|c|c|}
  \hline\hline
peak & HR & $\varphi\;(^\circ)$ & $|F_f||F_s|$ & peak & HR & $\varphi\;(^\circ)$ & $|F_f||F_s|$ \\
  \hline
1 & $4 \bar{2} 6_s+\bar{1} \bar{3}\,\bar{10}_f $ & 5.24 & 3.9 & 31  & $\bar{1} 1\bar{5}_f+0 2 2_s $ & 22.86 & 4.8 \\
2 & $2 0\,\bar{10}_f+4 4 0_s $ & 6.42 & 4.7 & 32 & --- & 26.94 & --- \\
3 & $0 1\,\bar{10}_f+2 4 2_s $ & 15.57 & 10.8 & 33 & $0 0 4_s+\bar{2} 0\bar{5}_f $ & 29.83 & 4.9 \\
4 & $\bar{1} 3\,\bar{10}_f+0 6 2_s $ & 17.54 & 8.6 & 34 & $4 0 0_s+2 \bar{2}\bar{5}_f $ & 30.10 & 31.4 \\
5 & $2 0 6_s+\bar{2} \bar{1}\,\bar{10}_f $ & 20.70 & 8.1 & 35 & --- & 33.03 & --- \\
6 & $4\bar{2} 6_s+\bar{1} \bar{3}. \bar{10}_f $ & 26.95 & 3.6 & 36 & $1 0 \bar{5}_f+2 2 0_s $ & 37.20 & 7.7 \\
7 & $6 \bar{2} 4_s+1 \bar{4}\,\bar{10}_f $ & 32.84 & 4.8 & 37 & $\bar{1} 1 \bar{5}_f+0 2 2_s $ & 22.66 & 6.0 \\
8 & $6 0 2_s+\bar{1} 0\,\bar{10}_f $ & 39.30 & 10.9 & 38 & $6 \bar{2} 0_s+3 \bar{4} \bar{5}_f $ & 38.39 & 6.3 \\
9 & $1 2\,\bar{10}_f+2 6 0_s $ & 42.49 & 26.4 & 39 & --- & 46.18 & --- \\
10 & $0 1\,\bar{10}_f+2 4 2_s $ & 44.36 & 23.6 & 40 & $1 3\, \bar{5}_f+0 6 \bar{2}_s $ & 53.77 & 4.0 \\
11 & $\bar{2} 2\,\bar{10}_f+0 4 4_s $ & 53.44 & 2.9 & 41 & $\bar{3} 2 \bar{5}_f+\bar{2} 2 4_s $ & 56.90 & 4.7 \\
12 & $6 \bar{2} 4_s+1 \bar{4}\,\bar{10}_f $ & 54.77 & 4.0 & 42 & $4 \bar{4} 4_s+0 \bar{4} \bar{5}_f $ & 59.83 & 7.5 \\
13 & $\bar{3} 4\,\bar{10}_f+\bar{2} 6 4_s $ & 65.10 & 5.6 & 43 & $\bar{4} 4 \bar{5}_f+\bar{4} 4 4_s $ & 60.24 & 2.5 \\
14 & $4 0 4_s+0 \bar{2}\,\bar{10}_f $ & 66.28 & 6.3 & 44 & $2 \bar{2} 4_s+\bar{1} \bar{2} \bar{5}_f $ & 63.13 & 8.7 \\
15 & $4 2 2_s+1 \bar{1}\,\bar{10}_f $ & 75.85 & 6.5 & 45 & $6 0 \bar{2}_s+4 \bar{3} \bar{5}_f $ & 66.20 & 12.3 \\
16 & $6 2 0_s+3 \bar{2}\,\bar{10}_f $ & 77.58 & 5.1 & 46 & --- & 73.82 & --- \\
17 & $\bar{1} 3\,\bar{10}_f+0 6 2_s $ & 80.71 & 12.9 & 47 & $4 2 \bar{2}_s+3 \bar{1} \bar{5}_f $ & 78.63 & 11.0 \\
18 & $\bar{3} 4\,\bar{10}_f+\bar{2} 6 4_s $ & 87.09 & 4.9 & 48 & $\bar{1} 4 \bar{5}_f+0 2 2_s $ & 81.54 & 6.1 \\
19 & $\bar{4} 3\,\bar{10}_f+\bar{2} 4 6_s $ & 92.98 & 3.6 & 49 & $2 2 0_s+1 0 \bar{5}_f $ & 82.73 & 16.9 \\
20 & $\bar{3} 1\,\bar{10}_f+0 2 6_s $ & 99.30 & 12.9 & 50 & --- & 86.97 & --- \\
21 & $6 0 2_s+2 \bar{3}\,\bar{10}_f $ & 102.56 & 3.7 & 51 & $0 2 \bar{5}_f+0 4 0_s $ & 89.90 & 13.7 \\
22 & $4 2 2_s+1 \bar{1}\,\bar{10}_f $ & 104.43 & 17.1 & 52 & $\bar{2} 0 \bar{5}_f+0 0 4_s $ & 90.17 & 4.8 \\
23 & $4 4 0_s+2 0\,\bar{10}_f $ & 113.72 & 6.3 & 53 & --- & 93.03 & --- \\
24 & $\bar{4} 3\,\bar{10}_f+\bar{2} 4 6_s $ & 114.97 & 13.5 & 54 & $2 0 2_s+0 \bar{1} \bar{5}_f $ & 97.20 & 7.0 \\
25 & $4 \bar{4} 4_s+0 \bar{4} \bar{5}_f $ & 0.17 & 7.5 & 55 & $\bar{4} 1 \bar{5}_f+\bar{2} 0 6_s $ & 98.39 & 11.7 \\
26 & $3 \bar{1} \bar{5}_f+4 2 \bar{2}_s $ & 3.17 & 12.1 & 56 & $4 \bar{2} 2_s+1 \bar{3} \bar{5}_f $ & 101.38 & 7.4 \\
27 & $\bar{1} 4 \bar{5}_f+\bar{2} 6 0_s $ & 6.16 & 7.0 & 57 & --- & 106.25 & --- \\
28 & --- & 17.54 & --- & 58 & $6 \bar{2} 0_s+3 \bar{4} \bar{5}_f $ & 113.77 & 5.8 \\
29 & $\bar{2} 3 \bar{5}_f+\bar{2} 4 2_s $ & 18.62 & 6.7 & 59 & $2 4 \bar{2}_s+2 1 \bar{5}_f $ & 116.83 & 7.0 \\
30 & $0 \bar{2} 6_s+\bar{3} \bar{1} \bar{5}_f $ & 21.61 & 6.1 & 60 & $\bar{4} 4 \bar{5}_f+\bar{4} 4 4_s $ & 119.83 & 3.8 \\
 \hline
  \hline
\end{tabular}}
\end{table*}

\section{refraction}

\begin{figure*}
  \includegraphics[width=5in]{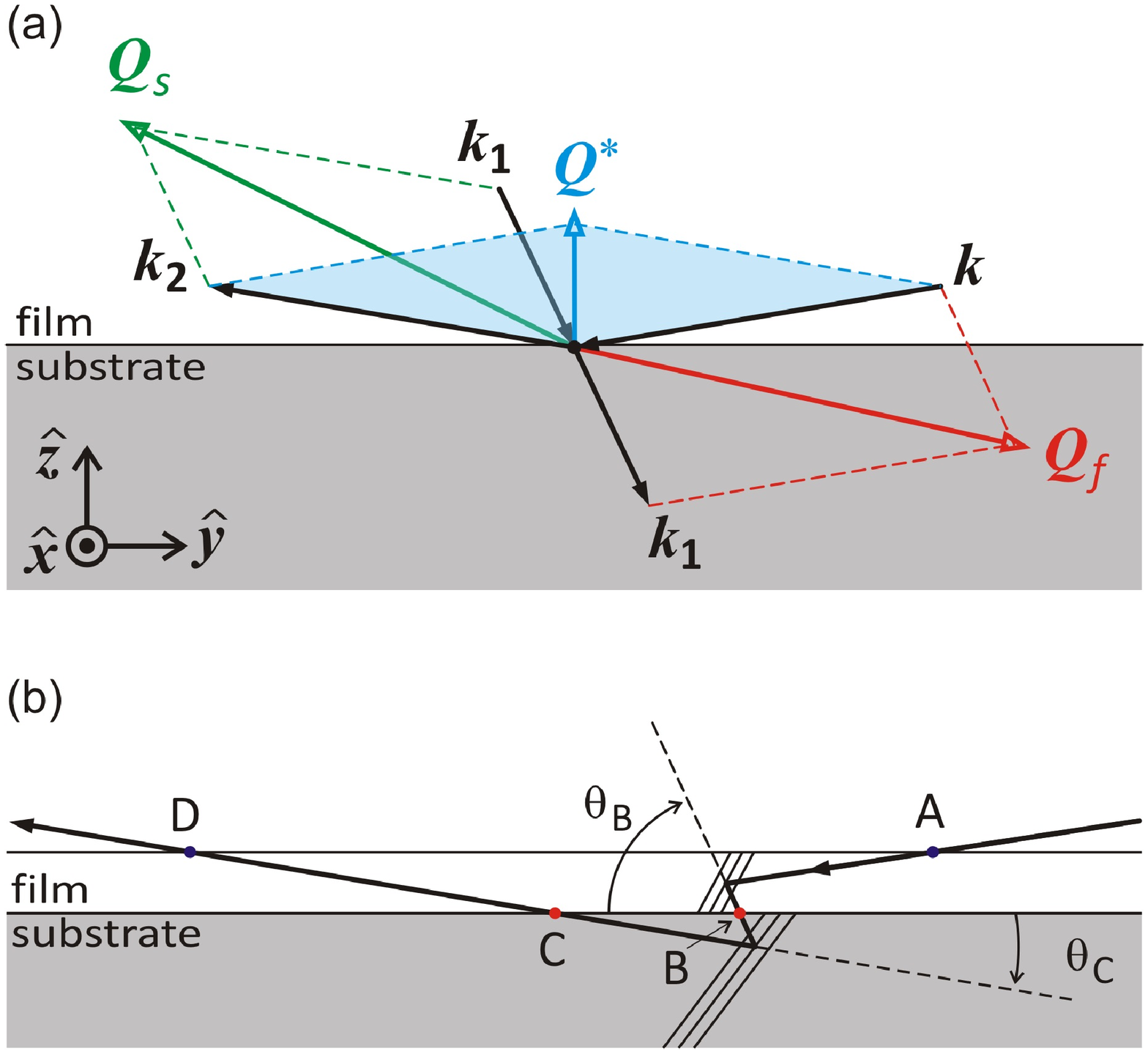}\\
  \caption{(a) Diffraction vectors and wavevectors of HR $\bar{4} 4 \bar{5}_f+\bar{4} 4 4_s $. $\vec{k}_1=\vec{Q}_f + \vec{k}$ and $\vec{k}_2=\vec{Q}_s + \vec{k}_1=\vec{Q}^* + \vec{k}$. (b) Scheme in real space of this HR.}\label{fig:hr6refraction}
\end{figure*}

Difference in film and substrate indexes of refraction for x-rays, $n_f$ and $n_s$ respectively, can shift in opposite sense the rocking curve peaks of hybrid pairs, even when there is no physical mismatch in the lateral lattice parameters. First, there is the change due to the optical lattice mismatch caused by the difference of wavelength in each material, resulting in $(\Delta a/a)_{\rm opt} = (n_f-n_s)/n_s$. Second, there is the angular deviation due to different angles in which the x-rays cross the film/substrate interface, as indicated in Fig.~\ref{fig:hr6refraction}. The angular deviations at the entrance and exit surfaces, points A and D in Fig.~\ref{fig:hr6refraction}(b), cancel each other. But, there are the deviations $\Delta\theta_{\rm B} = \arccos[(n_f/n_s)\cos\theta_{\rm B}]-\theta_{\rm B}$ and $\Delta\theta_{\rm C} = \arccos[(n_s/n_f)\cos\theta_{\rm C}]-\theta_{\rm C}$ at the common interface, points B and C in Fig.~\ref{fig:hr6refraction}(b). In a pair of HRs where the optical path of one HR is the time-reversed path of the other HR, the split in their rocking curve peaks is given by $2\Delta\theta_i=\Delta\theta_{\rm B}-\Delta\theta_{\rm C}$ as far as $n_s/n_f>1$.

For x-rays of 8\,keV, $n_f=1-1.9752\times10^{-5}$ (Bi$_2$Te$_3$) and $n_s=1-1.3063\times10^{-5}$ (BaF$_2$). Therefore, $(\Delta a/a)_{\rm opt} = -6.7\times10^{-6}$, and for the HR depicted in Fig.~\ref{fig:hr6refraction}, $\theta_{\rm B}=24.406^\circ$ and $\theta_{\rm C}=9.243^\circ$, leading to $2\Delta\theta_i=0.0032^\circ$. HR with first reflection in the film deflects towards higher angles, i.e. the contrary of that observed for HRs $\bar{4} 4 \bar{5}_f+\bar{4} 4 4_s $ and $4 \bar{4} 4_s+0 \bar{4} \bar{5}_f $ [Fig. 4(d-f), main text]. In the case of the other pairs in Fig. 4 (main text) with incidence angle of $\theta=18.738^\circ$, peak shiting due to refraction is even smaller, i.e. $2\Delta\theta_i<0.0015^\circ$.

Structure factor lists and refraction indexes have been calculated using the MatLab scripts \texttt{diffraction.m}, \texttt{sfactor.m}, \texttt{asfQ.m}, and  \texttt{fpfpp.m} available in open codes at the internet (back matter of https://link.springer.com/book/10.1007\%2F978-3-319-19554-4).

\end{document}